\documentclass[superscriptaddress,twocolumn,showpacs,preprintnumbers,amsmath,amssymb,pre,lettersize]{revtex4}

\usepackage{graphicx}
\usepackage{bbm}
\usepackage{subfigure}

\def\>{\rangle}
\def\<{\langle}

\def\Tr{{\text{Tr}}}
\def\ima{\imath}

\newcommand{\rmd}{{\rm d}}
\newcommand {\etal} {{\em et al}}

\begin{document}

\title{First verification of generic fidelity recovery in a dynamical system}

\author{Carlos Pineda}
\email{carlospgmat03@yahoo.com}
\affiliation{Instituto de  F\'{\i}sica, UNAM, Mexico}
\affiliation{Centro de Ciencias F\'{\i}sicas, UNAM, Mexico}
\author{Rudi Sch\"afer}
\affiliation{Fachbereich Physik der Philipps-Universit\"at Marburg,
D-35032 Marburg, Germany}
\author{Toma\v z Prosen}
\affiliation{Physics Department, Faculty of Mathematics and Physics, 
University of Ljubljana, Ljubljana, Slovenia}
\author{Thomas Seligman}
\affiliation{Centro de Ciencias F\'{\i}sicas, UNAM, Mexico}
\affiliation{Centro Internacional de Ciencias, Cuernavaca, Mexico}

\date{\today}
\begin{abstract}
  We study the time evolution of fidelity in a dynamical many body system,
  namely a kicked Ising model, modified to allow for a time reversal
  invariance breaking. We find good agreement with the random matrix
  predictions in the realm of strong perturbations. In particular for the
  time-reversal symmetry breaking case the predicted revival at Heisenberg
  time is clearly seen.
\end{abstract}


\maketitle

\section{Introduction}

The random matrix (RMT) description of fidelity decay~\cite{1367-2630-6-1-020}
has been successful in describing the behavior of dynamical models and actual
experiments~\cite{1367-2630-7-1-152,gorin-weaver}. While in this early work
linear response (LR) results were exponentiated (ELR) to describe long-time
behavior quite well, in a more recent paper it was shown, that the RMT model
presented in~\cite{1367-2630-6-1-020} can be solved exactly in the limit of
large Hilbert space dimension~\cite{shortRMT,1367-2630-6-1-199}.  These exact
results show a puzzling, though in absolute terms weak, revival of the
ensemble averaged fidelity amplitude at the Heisenberg time. The predicted
revival is strongest for the Gaussian symplectic ensemble (GSE), and weakest
for the Gaussian orthogonal ensemble (GOE). For the time being experiments are
far from the realm of parameters where this effect or any significant
deviations from ELR can be seen; neither have such deviations been seen in
dynamical models.

In the present paper we intend to show that this effect exists in a dynamical
many particle model in which we can carry calculations beyond the Heisenberg
time for very large systems, and for large perturbation strength. The system
we shall consider is a modification of the kicked Ising chain~\cite{prosenKI}.
Since we are dealing with a kicked system we have to focus on the Floquet
operator and hence we shall have to compare to the so called circular
ensembles of unitary matrices.  Note that in the large dimension limit the
statistics for spectral fluctuations coincide.  We introduce a multiple kick
in order to break an anti-unitary symmetry that plays the role of time
inversion and which we still call time reversal invariance (TRI).  To consider
TRI breaking evolution is useful as the revival at Heisenberg time is in
absolute terms three orders of magnitude larger for the Gaussian unitary
ensemble (GUE) than for the GOE.  The unperturbed Hamiltonians used for both
the TRI and the non-TRI situations are fixed such as to yield very good
agreement with the spectral statistics expected for the GOE or GUE,
respectively. This choice is certainly not unique as we could further modify
the kick-structure, but this is not contemplated.  The perturbation strength,
that enters the RMT model as a free parameter, can be determined from the
dynamics via correlation functions.  In both cases we shall thus find good
agreement between the fidelity of the dynamical model and RMT without any free
parameters.

The effects we look at are of the order of a part in a thousand or less for
the ensemble average, and we wish to analyze whether they can be seen in an
individual system.  For this to be the case we need large enough Hilbert
spaces such that the time evolution samples a large domain of
eigen-frequencies of the system, which in turn gives the opportunity for
self-averaging. In the specific case of kicked spin chains we can numerically
handle time evolution for up to 24 qubits for moderate times.

Fidelity describes the evolution of a cross-correlation function of a given
initial state under two different Hamiltonians or, equivalently, the evolution
of the auto-correlation function under the so-called {\em
  echo-dynamics}~\cite{reflosch}. The meaning of the latter term is a
composition of the forward time evolution by some Hamiltonian and backward
time evolution by a slightly different one.  For an initial state $|\psi(0)\>$
evolving under some \textit{unperturbed} unitary dynamics $U_0 (t)$ and the
same state evolving under a perturbed but still unitary dynamics
$U_\epsilon(t)$ the fidelity amplitude reads as
\begin{equation} \label{eq:firstfidelity} 
f(t) = \< \psi(0) |U_0^\dagger(t) U_\epsilon (t) |\psi(0)\>. 
\end{equation}
Fidelity  itself is defined as $F(t) = | f(t) |^2$ . 

We shall first review the RMT results, and give a brief resume of the linear
response and exact solutions. Next we discuss the generalized kicked spin
model and show that we can reach sufficiently large Hilbert spaces and
sufficiently strong perturbations to be in the desired regime. First we check
that for a suitable non-integrable parameter regime of the model we can have
GOE statistics for the TRI conserving model and GUE statistics for
the TRI violating model. Finally, we compare the model results for fidelity
decay with the corresponding RMT predictions.

\section{Random matrix results}

Let $H_0$ be the unperturbed Hamiltonian, chosen from one of the Gaussian
ensembles, and
\begin{equation}\label{def:h_epsilon} 
    H_\epsilon=H_0+\frac{\sqrt{\epsilon}}{2\pi} V 
\end{equation} 
the perturbed one. This somewhat unusual definition of the perturbation
strength $\epsilon$ has been applied for convenience of comparison with the
RMT results.  The fidelity amplitude is then given by
Eq.(\ref{eq:firstfidelity}) with $U(t)=\exp(-2\pi\imath H_0 t)$ and
$U_\epsilon(t)=\exp(-2\pi\imath H_\epsilon t)$.  It is assumed that $H_0$ has
mean level spacing of one, and thus $t$ is given in units of the Heisenberg
time $\tau_\textrm{H}$. The variance of the off-diagonal elements of $V$ is
chosen to be one.

Gorin \etal~\cite{1367-2630-6-1-020} calculated the Gaussian average of the
fidelity amplitude in the regime of small perturbations using the
linear-response approximation,
\begin{equation}\label{eq:linresp} 
  f(t)\sim 1- \epsilon \, C(t)\,. 
\end{equation} 
where $C(t)$ is given by 
\begin{equation}\label{eq:C_tau} 
    C(t)=\frac{t^2}{\beta}+\frac{t}{2}-\int_0^t \int_0^\tau 
b_{2,\beta}(\tau^\prime) \rmd \tau^\prime \rmd \tau \, , 
\end{equation} 
$1-b_{2,\beta}(t)$ is the spectral form factor, and $\beta$ is the
universality index, i.\,e. $\beta=1$ for GOE, $\beta=2$ for GUE, and $\beta=4$
for GSE. For an explicit calculation, knowledge of the spectral form factor is
thus sufficient. Using the ELR approximation,
\begin{equation}\label{eq:explinresp} 
  f(t)\sim e^{-\epsilon \, C(t)}\,, 
\end{equation} 
the authors were able to describe quantitatively the cross-over from Gaussian
to exponential decay with increasing perturbation strength.
 
It is obvious that the linear-response approximation must break down for large
perturbations.  In a recent paper St\"ockmann and Sch\"afer applied
super-symmetry techniques to calculate Gaussian averages of the fidelity decay
for arbitrary perturbation strengths~\cite{1367-2630-6-1-199}.  In order to
apply the super-symmetry framework, the authors considered the {\em state
  averaged} version of Eq.(\ref{eq:firstfidelity}) using the Hamiltonians
define above and the perturbation parameter $\epsilon$
\begin{equation}\label{eq:fidelitytrace}
  f(t)=\frac{1}{N}{\rm Tr}\left[e^{2\pi\imath H_\epsilon t}
  e^{-2\pi\imath H_0 t}\right]\,,
\end{equation}
where $N$ denotes Hilbert space dimension, hence making it independent of the
state considered.

The exact results obtained in Ref.~\cite{1367-2630-6-1-199} are in good
agreement with the linear-response result for very small perturbations and
also confirm the validity of the ELR approximation for moderate perturbation
strengths, and indeed whenever the fidelity is not too small.  However, these
calculations also revealed an important generic feature, which motivated the
present work: for very strong perturbations, the fidelity amplitude shows a
partial recovery at the Heisenberg time.

For the GUE case the result for the fidelity amplitude reads
\begin{equation}\label{41} 
  f(t)=\left\{\begin{array}{ll} 
    e^{-\frac{\epsilon}{2}t}\left[s(\frac{\epsilon}{2}t^2) 
    -t s'(\frac{\epsilon}{2}t^2)\right]\,,\qquad &t\le1 \\ 
    e^{-\frac{\epsilon}{2}t^2}\left[s(\epsilon t) 
    -\frac{1}{t}s'(\frac{\epsilon}{2}t)\right]\,,&t> 
    1 
  \end{array}\right.\,, 
\end{equation} 
where
\begin{equation}\label{42} 
  s(x)=\frac{\sinh(x)}{x}\, , 
\end{equation} 
and $s'(x)$ denotes its derivative.  For the GOE case, the result is not so
simple and is expressed in terms of one-dimensional
integrals~\cite{1367-2630-6-1-199}.

\section{The multiply kicked Ising model}

\begin{figure}
     \includegraphics{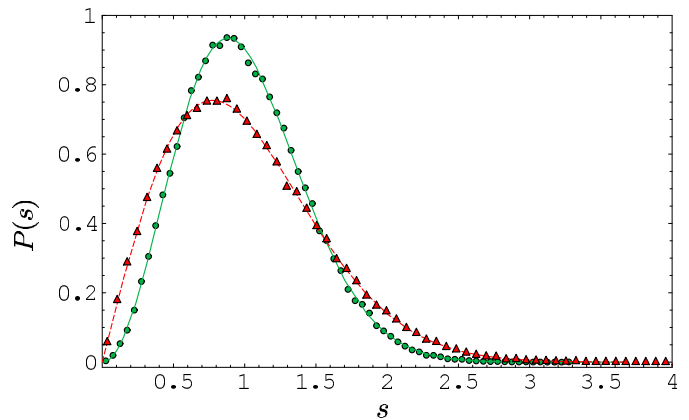}
     \caption{We show the probability density for the nearest
       neighbor spacing for 18 qubits and averaged over the different relevant
       $\mathcal H_{k}$ spaces ($k=1,\dots,8$).  Numerical results are shown
       in green circles for the model corresponding to the GUE and
       red triangles for the GOE.  Parameters are $M=2$,
       $b^{(1)}=(1,1,0)$, and $ b^{(2)}=(1.4,0,1.4)$ for the GUE case and $
       b^{(1)}=(1.4,1.4,0)$, for the GOE case.  We plot the exact RMT result
       for both the GOE (dashed red line) and the GUE 
       (continuous green line) since comparison with Wigner surmise is not
       satisfactory.}
  \label{fig:Ps}
\end{figure}

We next introduce the multiply kicked Ising model (MKI) as a modification of 
the kicked Ising model introduced in~\cite{prosenKI}. The evolution operator
corresponding to one time step is
\begin{equation}\label{eq:mki}  
U_\text{MKI} = \prod_{n=1}^{M} U_\text{Ising}U_\text{kick}^{(n)}
\end{equation}
with
\begin{align}
U_\text{Ising}& =\exp \left( -\ima J \sum_{j=0}^{L-1} \sigma ^z_j \sigma^z_{j+1} \right),\\
U_\text{kick}^{(n)}&=\exp\left(-\ima  \sum_{j=0}^{L-1}    \vec{b}^{(n)}  \cdot\vec{\sigma}_j \right),
\label{eq:hamiltoniangenerla}
\end{align}
and periodic boundary conditions $\sigma_{L} \equiv \sigma_0$.  Here $\sigma
^{x,y,z}_j$ are the Pauli matrices corresponding to spin 1/2 at position $j$,
and $\vec{\sigma}_j=(\sigma^x_j,\sigma^y_j,\sigma ^z_j)$. The system defined
in Eq.(\ref{eq:mki}) represents a periodic 1-d array of $L$ spin $1/2$
particles which periodically receive a sequence of $M$ different kicks of
instantaneous magnetic field pulses equally spaced in time.  The free
evolution ($U_\text{Ising}$) is simply given by the Ising interaction between
nearest neighbors, with dimensionless strength $J$ which we fix to 1 in the
numerical simulations.  Each of the $M$ kicks is an instantaneous and
homogeneous magnetic field characterized by the dimensionless vector
$\vec{b}^{(n)}$.  Due to the two body interaction of the system, we are able
to evaluate efficiently the evolution of any state~\cite{prosenKI}, and hence,
as noted in~\cite{effidiag}, also its spectrum.  In the previous
model~\cite{prosenKI} the same kick was applied periodically, that is $M=1$.

\begin{figure}
     \includegraphics{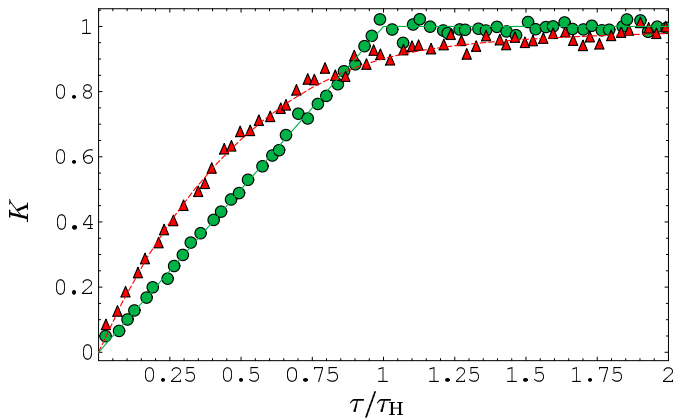}
     \caption{Same parameters and sign conventions as in Fig.~\ref{fig:Ps}. 
       We show the spectral form factor averaged over all relevant $\mathcal
       H_k$'s.  In order to smooth the function, we also average over a window
       in $t$ of size $0.15 t_\text{H}$.}
  \label{fig:K2}
\end{figure}

Since we wish to compare the behavior of fidelity in our model-system with the
average behavior of an ensemble of random matrices, it is necessary to
identify the correct symmetry class.  To this end, we have to study the
symmetries of the MKI model.  As couplings between neighboring qubits as well
as the effect  of the kick on all qubits  are equal, it is easy  to see that we
formally have a ring of qubits and thus a rotational as well as a reflection symmetry.

Let us define the rotational operator $\hat{R}$ over the elements of the
computational basis, as $ \hat{R} |i_0 i_1 \dots i_{L-1}\> = |i_{L-1} i_0
\dots i_{L-2}\> $, where each $i_j\in \{0,1\}$, and extend it to the other
elements of the Hilbert space, requiring linearity.  If we visualize our model
as a ring, the action of this operator is to rotate the particles in the ring
by one position. This operator defines a rotation group ${\cal C}_L$.  We can
further define the reflection operator $\hat{P}$ on the computational basis as
$ \hat{P} |i_0 i_1 \dots i_{L-1}\> = |i_{L-1} i_{L-2} \dots i_{0}\>$ and again
extend it to the whole space requiring linearity.  The two operators do not
commute and form the group ${\cal C}_{L,v}$ of rotations and reflections of a
ring of $L$ elements.  Note that the $[U_{\textrm MKI},\hat P]=[U_{\textrm
  MKI},\hat R]=0$, hence the eigenvalues, $\exp(2\pi\imath k/L)$ with $k\in
\mathbb{Z}/L$, of the rotation operator define the invariant subspaces which
are degenerate with two parities except for the cases $k=0$ and $k=L/2$ the
latter for even $L$ only.  In the latter cases invariant subspaces of well
defined parity can be defined depending on the behavior under reflections,
though for a given set of qubits not necessarily both parities need occur. The
dimension for any set with fixed $k$ is approximately equal to $2^L/L$.

Since we do not identify other symmetries in the system, we expect to find a
range of parameters in which each of the blocks of $U_\text{MKI}$
corresponding to the minimal invariant subspaces, behave as a typical member of the
GUE.  This was numerically checked by analyzing the spectra in each
of the invariant subspaces, for the parameters shown in the caption of
Fig.~\ref{fig:Ps}.  The resulting nearest neighbor statistics and spectral
form factors (both defined in~\cite{mehta}) are presented in Fig.~\ref{fig:Ps}
and Fig.~\ref{fig:K2}, respectively.  Other statistical tests like the number
variance, skewness and excess were also applied with excellent agreement with
the expected behavior (not shown). The success of these tests is a strong
indicator that indeed no other symmetries are present.

Consider now the case in which we have only one periodic kick, that is $M=1$
in $U_\text{MKI}$.  Rotating each individual spin, the magnetic field
($\vec{b}^{(1)}$) can be made to have only components in the $xz$ plane.  The
Hamiltonian will have only real components in the basis in which
$\sigma^{x,z}_j$ are real, hence an anti-unitary symmetry becomes evident.
This symmetry $\mathcal{K}$ (complex conjugation in the basis mentioned above)
is \textit{not} time reversal, since spin is being
reflected in the $xz$ plane of each qubit, instead of being reversed. Furthermore, this symmetry will
change the sign of $k$ and hence, by itself, will not provide an anti-unitary
symmetry within the invariant subspaces.  Recalling that $\hat P$ also
reverses the sign of $k$, we observe that the combined operator $\hat
P\mathcal{K}$, will provide each of the invariant subspaces with an
anti-unitary symmetry, which we call time reversal invariance (TRI).  We
expect then that for $M=1$ the system will have a range of parameters in which
each invariant subspace will behave as a typical member of the GOE 
for some values of the parameters which are sufficiently well separated from
the exactly solvable cases of longitudinal and transverse
fields~\cite{prosenIntegrableChaos}.  Numerical evidence favoring this
statement is presented in Figs.~\ref{fig:Ps}~and~\ref{fig:K2}.

Effectively we will have, for each of the invariant subspaces a Heisenberg
time given by $t_\text{H} \approx 2^L/L$ (the approximate number of levels)
except for TRI subspaces for which we shall have $t_\text{H} \approx
2^{L-1}/L$.

%%%%%%%%%%%%%%%%%%%      
The perturbation  we shall consider  is a kick  in the $x$ direction:
\begin{equation}
A= \sum_{j=0}^{L-1} \sigma^x_j
\label{eq:perturbation}
\end{equation}
in the Floquet propagator [Eq.(\ref{eq:mki})]
\begin{equation}
U_{{\rm MKI},\delta} = U_{\rm MKI}\exp(-\imath \delta A).
\end{equation}
Comparing the linear response formula for the dynamical model~\cite{prosenKI},
\begin{equation}
f(t) = 1 - \frac{\delta^2}{2} \sum_{t'=-t+1}^{t-1} C(t') \;,
\end{equation}
with the correlation function $C(t) = 2^{-L}\Tr A U^\dagger(t) A U(t)$ with
the random matrix model, we connect the perturbation strengths
\begin{equation}\label{eq:relepsilondelta}
\epsilon = 2^L \delta^2 \sigma \;.
\end{equation}
Here
$\sigma = \lim_{t\to\infty}\lim_{L\to\infty}\frac{1}{2}\sum_{t'=-t}^t C(t')$
is the integrated correlation function of the unperturbed dynamics.

Note that the perturbation will have all the symmetries presented, so that it
will not mix the different invariant subspaces. Thus, if we consider an
initial state with components from all the $\mathcal H_k$ spaces, we will
obtain an average of fidelity over $L$ different initial conditions each in a
different $\mathcal H_k$ space.

\section{Results}

Since the effect we want to observe is extremely small, we must have a full
understanding of the most important causes of deviation in the evaluation of
fidelity in Eq.(\ref{eq:fidelitytrace}). Note that here we are taking two
averages, one of them over the ensemble of Hamiltonians and the other one over
the ensemble of initial conditions. The first of this averages we are {\em
  not} going to evaluate in our numerical model since we want to show that an
individual quantum chaotic system (in the limit of large Hilbert space
dimension) actually behaves according to the random matrix ensemble average
over the appropriate symmetry class.  The other average (over the Hilbert
space) is possible to evaluate in an exact way, taking the average over an
orthonormal basis resulting in a trace operation. In practice, for very large
Hilbert spaces this method is not efficient. Instead we shall use an approximation 
which introduces an error that can be made as small as desired.

\begin{figure}
\includegraphics{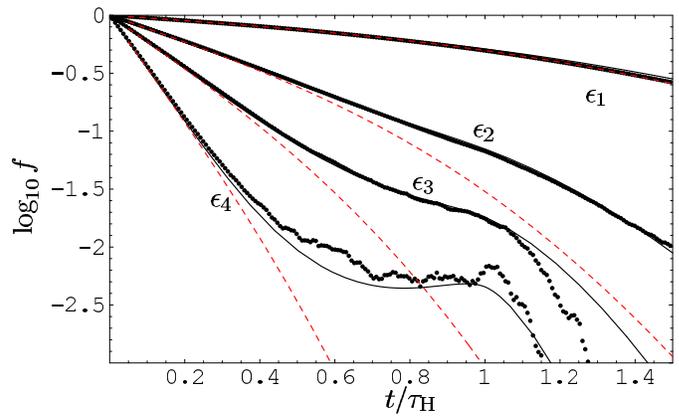}
\caption{Different perturbations for the GUE system. We have here 16 qubits
  and 10 initial conditions.  in Fig.~\ref{fig:Revival}. Dots show the
  numerical calculation (real part) and thin curves show RMT prediction
  whereas red dashed curves show the ELR approximation.  We set perturbations
  to $\epsilon_1=5.15$ (in this case all the three curves are overlapping),
  $\epsilon_2=10.3$, $\epsilon_3=15.455$, $\epsilon_4=20.6$. }
\label{fig:varye}
\end{figure}

Let us first comment on the latter source of error.  For any operator
$A$, we have that
\begin{equation}
\Tr A=\left< A \right>_\psi \;,
\end{equation}
where $\left< \cdot \right>_\psi$ denotes the average over Gaussian random
states: $|\psi\>=\sum_{i=1}^N x_i |i\>$, where $\{ |i\> \}_{i=1,\dots,N}$ is an
orthonormal basis and $x_i$ are independent complex Gaussian distributed
random numbers with standard deviation $1/\sqrt{N}$. Then, we have that
\begin{equation}\label{eq:traceapproximation}
\Tr A\approx \frac{1}{m} \sum_{j=1}^m \left<\psi_j | A |\psi_j\right>.
\end{equation}
Of course letting $m \to \infty$ will make the average exact. But for large
$N$ it turns out that already taking a small number of states is enough to
have a very good approximation of the expected value of $A$. So we shall
evaluate Eq.~(\ref{eq:fidelitytrace}) with the aid of
Eq.~(\ref{eq:traceapproximation}).  In particular for the sizes of the Hilbert
spaces considered here (upto $2^{20}$) the number of states needed to achieve
the precision required is much smaller than the dimension of the space. In our
case we want to evaluate the expectation value of the echo operator
\begin{equation}
M_t=U_\delta^\dagger(t) U(t)
\end{equation}
for different values of $t$. For a fixed value of $t$ we have a distribution
of values of $\left<\psi | M_t |\psi \right>$ and hence an associated standard
deviation. Although at $t=0$ this standard deviation is zero, after a
transient time it approaches an stationary value, determined by finite size
effects. We shall call this value $\sigma_\textrm{finite average}$. Then the
error in the evaluation of $\Tr M_t$ will be $\sigma_\textrm{finite
  average}/\sqrt{m}$ if we take $m$ sample states.  Estimating numerically
this value is trivial since for $m$ different realizations we can evaluate the
standard deviation at each time and then average over time, for $t$'s larger
than the transient time.

Considering only one particular Hamiltonian will also cause some deviations
from the exact RMT formula: recall that to obtain, say Eq.(\ref{41}), we
averaged over an ensemble of Hamiltonians.  Even if we obtain the exact trace
of the echo operator its value will fluctuate in time around the ensemble
averaged value. These fluctuations can be characterized with a standard
deviation $\sigma_\textrm{intrinsic}$.  For a fixed Hamiltonian the only way
to decrease this number is to increase the dimension of the system. In
practice, for computational reasons, we can only achieve results for up to 20
qubits for times of the order of Heisenberg time.

\begin{figure}
\includegraphics{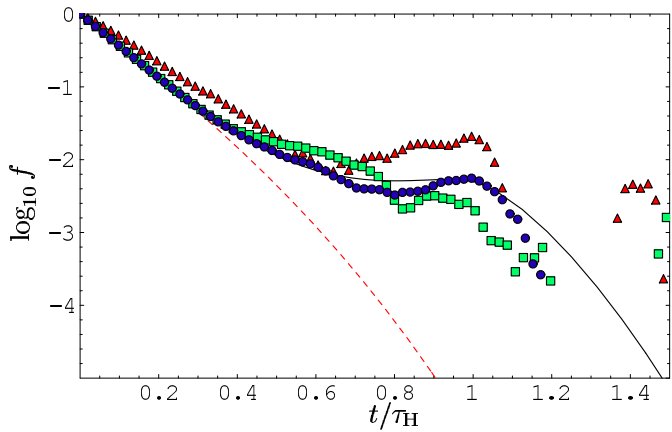}
\caption{Different number of qubits for a same effective perturbation 
  for the GUE system. The perturbation strength is $\epsilon=20$.  We vary the
  number of qubits, red triangles being 10, green squares 12, and blue circles
  16.  We have, in all cases, 200 initial conditions. We only show the points
  where $\textrm{Re} f>0$.  The red dashed line is the ELR prediction.}
\label{fig:varyq}
\end{figure}

Since those two effects can be assumed to be completely independent, the total
average deviation of Eq.(\ref{eq:fidelitytrace}) will be given by
\begin{equation}
\sigma_\textrm{total}^2=\sigma_\textrm{intrinsic}^2+\frac{\sigma_\textrm{finite average}^2}{m}.
\end{equation}
We can estimate a posteriori this quantity very easily, since we know that the
imaginary part of the ensemble averaged fidelity amplitude is zero for all
times: $\textrm{Im} f(t)=0$. Hence computing the fluctuations of the
imaginary part of the data obtained will give us $\sigma_\textrm{total}$ and
hence $\sigma_\textrm{intrinsic}$ (recall that evaluating
$\sigma_\textrm{finite average}$ is trivial).  The knowledge of
$\sigma_\textrm{intrinsic}$ will give us an estimate for the minimal value of
$m$ needed to estimate the value of $\Tr M_t$ with sufficient accuracy. In the
results shown in this section, we always have $\sigma_\textrm{finite
  average}/\sqrt{m} \ll \sigma_\textrm{intrinsic}$.

We now turn our attention to the numerical results obtained.  We shall set
$\vec{b}^{(1)}=(1.4, 1.4,0)$. For the GOE case ($M=1$) we do not need to
specify any other parameter except for the number of qubits and the strength
of the perturbation.  The resulting value of the integrated correlation
function is found to be $\sigma\approx 1.27$.
In case we want to observe the GUE prediction we set $M=2$, take the same
$\vec{b}^{(1)}$ as in GOE case and $\vec{b}^{(2)}=(1, 0,1)$. The integrated
correlation function is found to be $\sigma \approx 0.93$.

\begin{figure}
\includegraphics{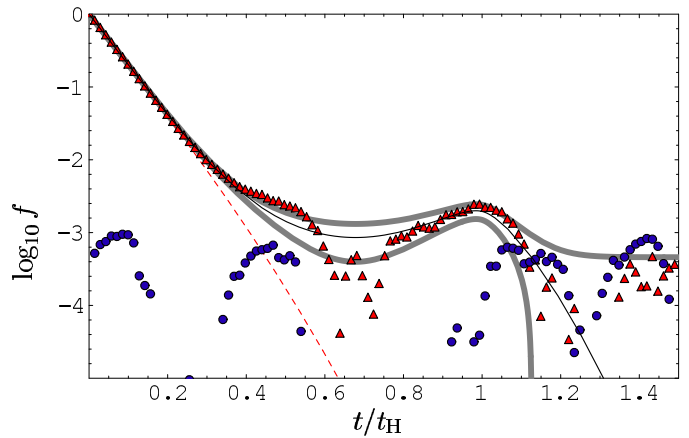}
\caption{Revival for 20 qubits.
  The perturbation strength is $\epsilon=31.78$. The red
  triangles indicate the real part, whereas the blue circles indicate the
  imaginary part (whenever they are greater than 0), in time steps of 525,
  calculated with 15 initial conditions.  The black thin curve is the RMT
  theoretical prediction, the red dashed line is the ELR prediction and the
  gray curves indicate the RMT curve plus/minus the calculated intrinsic
  deviation.}
\label{fig:Revival}
\end{figure}

In Fig.~\ref{fig:varye} we show, for the GUE-type system, how varying the
perturbation parameter changes the behavior of fidelity. Even for fairly
large perturbations ($\epsilon_1=5.15$) we obtain a fairly good approximation
with ELR, which almost coincides with the exact RMT result.  For larger
perturbations, deviation of the ELR from exact RMT are big enough to be
observed, and indeed they are observed clearly for $\epsilon_1=10.3$.  For
$\epsilon_4=20.6$, where we should observe a recovery, the behavior at those
low fidelities is shadowed by the intrinsic error.  Here we chose sufficiently
many initial conditions, so that the finite average error is considerably
smaller than the intrinsic error.

In Fig.~\ref{fig:varyq} we show, again for the GUE-type system, how increasing
the number of qubits decreases the deviation from RMT prediction. We fix the
effective perturbation [which scales with the number of qubits as shown in
Eq.(\ref{eq:relepsilondelta})] and observe the behavior of the real part of
fidelity for 10, 12 and 16 qubits.  Particularly for 10 qubits we see a
noticeable deviation, probably due to the non-vanishing correlation function
for large times. This value becomes smaller as the number of qubits increases
but also we average over a larger number of invariant subspaces. In all cases
we can see that the correspondence with exact RMT is much better than with
ELR.

In Fig.~\ref{fig:Revival} we show the evolution of fidelity, for $\epsilon=
31.78 $ where the revival is strongest.  The error due to finite averaging is
much smaller than the intrinsic error. We cannot diminish intrinsic error
further, since for 20 qubits we need near more than two weeks of computer time
per initial condition, and in order to increase the system size by one qubit
we need 4 times more CPU time for each initial condition, since for each time
step we roughly need to double the number of operations and we also need to
double the number of time steps (due to the increase in the Heisenberg time).
We expect that for sufficiently large Hilbert spaces the intrinsic error will
be so small and the self averaging so strong that even for one initial
condition one can clearly observe the fidelity revival.

\begin{figure}
\includegraphics{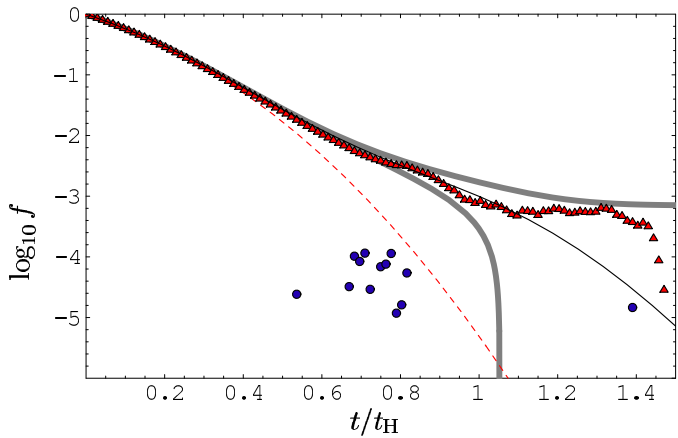}
\caption{We show that our GOE system indeed behaves as given by the RMT theory.
  Same color coding as in Fig.~\ref{fig:Revival}.  The perturbation strength
  is $\epsilon=10$, and the number of initial conditions is 15.  The size of
  the system is 20 qubits, $h_\perp=h_\parallel=1.4$.  We show only the points
  where $f>0$. The red dashed line is the ELR prediction.}
\label{fig:GOE}
\end{figure}

Finally, for completeness, we also show the behavior for the GOE-type system
in Fig.~\ref{fig:GOE}.  The deviation with ELR is evident, and the error is
within the bounds.

\section{Conclusions}

The recent exact solutions of the random matrix model for fidelity decay
display quite unexpected characteristics for large perturbations and large
times. The most remarkable one is the maximum that develops at Heisenberg
time. We showed that this phenomenon, established for an ensemble average, can
actually be seen in individual dynamical systems if the Hilbert space is large
enough: the effect is not overshadowed by noise and finite size effects.
Self-averaging is effective, and indeed we find good agreement with RMT
results for fidelity decay for two individual kicked spin chain models, one
with, one without a \textit{pseudo} time reversal invariance.  The results are
achieved without any fit parameter as the perturbation strength has been
determined directly from the model we use.

The model used was a kicked spin chain which can reproduce both TRI conserving
and TRI breaking dynamics. It has no obvious classical analogue, so we could
feel reasonably confident that it would display generic behavior in a regime
far from the perturbative one.  Whether the maximum at Heisenberg time can
also be expected for systems with a classical analog is an open and
interesting question. Clearly, for strong perturbations and short times they
should not follow RMT behavior, but that does not exclude, that this
behavior is recovered at least qualitatively for times as long as the
Heisenberg time.

\begin{acknowledgments}
  We acknowledge support from DGAPA-UNAM project IN101603 and CONACyT, Mexico,
  project 41000 F.  The work of C.P. was supported by Direcci\'on General de
  Estudios de Posgrado (DGEP).  R.S. acknowledges support by the Deutsche
  Forschungsgemeinschaft.  T.P. wishes to thank CiC (Cuernavaca) where major
  parts of this work have been performed for hospitality and acknowledges
  support from Slovenian Research Agency (programme P1-0044 and grant
  J1-7347).
\end{acknowledgments}

\bibliographystyle{apsrev}
\bibliography{paperdef,paper,book,miblibliografia}
\end{document}